# Chaotic time-delay signature suppression using quantum noise

YANQIANG GUO[1,2], XIN FANG[1], HAOJIE ZHANG[1], TONG ZHAO[1], MARTIN VIRTE[3], AND XIAOMIN GUO[1,2,*]

[1]Key Laboratory of Advanced Transducers and Intelligent Control System, Ministry of Education, College of Physics and Optoelectronics, Taiyuan University of Technology, Taiyuan 030024, China
[2]State Key Laboratory of Cryptology, Beijing 100878, China
[3]Brussels Photonics, Department of Applied Physics and Photonics, Vrije Universiteit Brussel, Brussels 1050, Belgium
*Corresponding author: guoxiaomin@tyut.edu.cn

**Time-delay signature (TDS) suppression of semiconductor lasers with external optical feedback is necessary to ensure security of chaos-based secure communications. Here we numerically and experimentally demonstrate a technique to effectively suppress the TDS of chaotic laser using quantum noise. The TDS and dynamical complexity are quantified using the autocorrelation function and normalized permutation entropy at the feedback delay time, respectively. Quantum noise from quadrature fluctuations of vacuum state is prepared through balanced homodyne measurement. The effects of strength and bandwidth of quantum noise on chaotic TDS suppression and complexity enhancement are investigated numerically and experimentally. Compared to the original dynamics, the TDS of this quantum-noise improved chaos is sup-pressed up to 94% and the bandwidth suppression ratio of quantum noise to chaotic laser is 1:25. The experiment agrees well with the theory. The improved chaotic laser is potentially beneficial to chaos-based random number generation and secure communication.**

Semiconductor lasers subject to external optical feedback (EOF) represent a prominent platform for studying complex dynamics [1]. Due to instability-induced large intensity, wide-band and high-dimensional "noise" [2-5], the optical-feedback laser chaos has inspired a number of emerging applications, such as chaos-based secure communications [6-8], chaos key distribution [9], photonic information processing [10], physical random number generation (RNG) [11-13], chaotic optical sensing [14], chaotic lidar [15]. However, the EOF gives rise to an inherent time-delay signature (TDS), which is identified using autocorrelation function (ACF) at the EOF delay time [16,17]. The TDS is almost inverse to the complexity of chaos, and the complexity can be quantified by permutation entropy (PE) [18]. The TDS could expose the external cavity length and degrade the complexity of chaos, deteriorating the quality of secure communication and randomness in RNG. It is essential to suppress the TDS and its reduction is beneficial to practical applications of chaotic lasers.

Up to now, many schemes have been proposed to suppress the TDS of chaotic lasers, including external parameter-condition tuning and post-processing, such as double-mirror feedback [19], phase modulation feedback [20,21], cascaded injection or coupling [22,23], fiber-Bragg-grating feedback [24,25], random distributed feedback [26], fiber propagation [27], exclusive-OR (XOR) operation [11], the m-least significant bits selection [12], and electrical heterodyne [28]. The above methods are also to extract random or stochastic noise from the chaotic process, and the stochastic spiking outputs of chaotic lasers are derived from intrinsic quantum noise. Which is a totally random and independent white noise arising from quantum fluctuations. While the nonlinear amplification of intrinsic quantum noise is entirely distinct from fluctuation induced by deterministic chaos, it is important to reveal the origins of the stochastic effects and extract random noise efficiently [29-31]. However, the effect of quantum noise on the underlying dynamics and randomness of chaotic laser remains to be explored.

In this letter, the chaotic TDS suppression using quantum noise is investigated numerically and experimentally. Quantum shot noise from measured quadrature fluctuations of vacuum state is extracted by balanced homodyne detection. The quantum noise provides non-deterministic initial values for the dynamical process of chaotic laser, and we numerically and experimentally study the influences of quantum noise strength and bandwidth on the TDS suppression and complexity enhancement of chaotic laser. Narrow bandwidth quantum noise substantially suppresses the TDS of wide bandwidth chaotic laser, and the observation is also confirmed using Lang-Kobayashi laser equations with stochastic Gaussian white noise.

The schematic illustration of the experimental setup is shown in Fig. 1. A laser source (Laser) with a wavelength of 1550 nm outputs a single-mode continuous laser beam, which acts as the local oscillator (LO). One group of half-wave plate (HWP) and polarizing beam splitter (PBS) ensures the interference of the LO and vacuum state. Another group of HWP and PBS serves as accurate beam splitting with two balanced outputs, and the transmitted and reflected outputs are coupled separately into two single-mode optical fibers by two triplet-lens couplers. Balanced homodyne detector (BHD, Thorlabs PDB480C) is used to acquire the quadrature fluctuations of the vacuum state and amplify the quantum shot noise to the macro level via LO and electrical gains. The resulting quantum shot noise is down-mixed using a radio frequency and filtered by different bandwidth low-pass filters. The output electrical signal is accurately controlled by variable wideband amplifier, and subsequently injected into a laser diode (LD, Eblana Photonics EP1550-0-DM-B05-FM) subject to EOF. The LD operates at 1550 nm and its threshold current $J_{th}$ is 10.3 mA. A temperature controller (TC) and a current source (CS) stabilize the LD with accuracies of 0.01°C and 0.1 mA, respectively.

The EOF is formed by an optical circulator (OC), a fiber coupler (FC) and a variable optical attenuator (VOA), and the EOF delay time is 86.7 ns. The feedback intensity and polarization are adjusted by the VOA and the polarization controller (PC). The outcome of the chaotic laser is fed into a 50 GHz photodetector (PD, Finisar XPDV212ORA-VF-FP), and the power spectrum and time series of chaotic signals are recorded by a 26.5 GHz RF spectrum analyzer (SA, Agilent N9020A, 3MHz RBW, 3KHz VBW) and an 80 GS/s real-time oscilloscope (OSC, Lecroy, LabMaster10-36Zi) with 36 GHz bandwidth, The sampling rate of the OSC is set at 10 GS/s with 1M recorded data for each sequence sample, that is, the time interval of each sequence sample is 100 μs. It's worth noting that the wavelength of the LO for quantum noise generation is optional and independent of that of the chaotic laser.

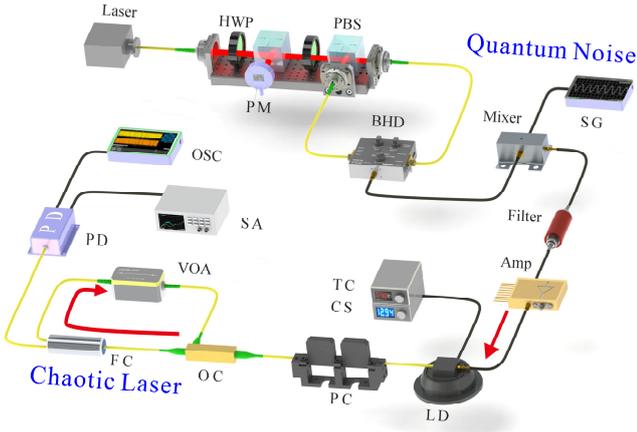

**FIG.1.** Schematic illustration of experimental setup. HWP: half-wave plate; PBS: polarizing beam splitter; PM: Power Meter; BHD: balanced homodyne detector; SG: signal generator; Amp: variable wideband amplifier; TC: temperature controller; CS: current source; LD: laser diode; PC: polarization controller; OC: optical circulator; FC: fiber coupler; VOA: variable optical attenuator; PD: photodetector; OSC: oscilloscope; SA: spectrum analyzer.

To investigate the influence of quantum noise on the dynamics of chaotic laser, the theoretical model can be described by the following Lang-Kobayashi rate equations with stochastic Gaussian white noise:

$$\dot{R}(t) = \frac{1}{2}[G(t) - \tau_p^{-1}]R(t) + \kappa R(t - \tau_{ext})\cos[\phi(t)] + \xi(t), \quad (1)$$

$$\dot{\varphi}(t) = \frac{\alpha}{2}[G(t) - \tau_p^{-1}] - \kappa R(t - \tau_{ext})R(t)^{-1}\sin[\phi(t)] + \eta(t), \quad (2)$$

$$\dot{N}(t) = \frac{J}{e} - \frac{N(t)}{\tau_N} - G(t)|R(t)|^2 + \beta(t), \quad (3)$$

$$\phi(t) = \omega\tau + \varphi(t) - \varphi(t - \tau_{ext}). \quad (4)$$

In the equations, the updated Gaussian white noise terms $\xi(t)$, $\eta(t)$ and $\beta(t)$ are added in the electric amplitude, phase and carrier density equations. The $\xi(t)$, $\eta(t)$ and $\beta(t)$ represent uncorrelated white Gaussian noise with zero mean and $<\xi(t)\xi(t-\tau_{ext})> = D\delta(\tau_{ext})$, $<\eta(t)\eta(t-\tau_{ext})> = D\delta(\tau_{ext})$, $<\beta(t)\beta(t-\tau_{ext})> = D\delta(\tau_{ext})$, where $D$ is the noise strength. $R(t)$ is the optical field amplitude, $\varphi(t)$ is the field phase, and $N(t)$ is the carrier density. The nonlinear optical gain $G(t)$ is given by $G(t) = G_N[N(t)-N_0]/(1+\varepsilon|R(t)|^2)$, $N_0$ is the carrier density at transparency, $\alpha$ is the line-width enhanced factor, $\tau_p$ and $\tau_N$ are photon lifetime and carrier lifetime, $\kappa = (1-r^2_{in})r_0/(r_{in}\tau_{in})$ is the optical feedback strength ($r_{in}$ is the reflectivity of the internal cavity, $r_0$ is the reflectivity of the external mirror, and $\tau_{in}$ is the optical round-trip time in internal cavity), $\tau_{ext} = 2L/c$ is the EOF delay time ($L$ is the length of the external cavity and $c$ is the speed of light in the medium), $\omega = 2\pi c/\lambda$ is the angular optical frequency ($\lambda$ is the optical wavelength), $e$ is the unit charge, and $J=\rho J_{th}$ is the injection current density ($J_{th}$ is the laser threshold current and $\rho$ is the pump factor). The following parameter values are set according to the experimental operation: $\alpha$=5, $G_N$=2.56×10$^{-8}$$ps^{-1}$, $N_0$=1.35×10$^8$, $\tau_{ext}$=86.7ns, $\tau_p$=3.2ps, $\tau_N$=2.3ns, $\lambda$=1.55μm, $\varepsilon$=5×10$^{-7}$.

ACF at the EOF delay time is used to quantify the TDS of chaotic laser, and the measure between a signal and its time-delay version is defined as follows:

$$C(\Delta t) = \frac{\left\langle\left[I(t+\Delta t) - \langle I(t+\Delta t)\rangle\right]\left[I(t) - \langle I(t)\rangle\right]\right\rangle}{\sqrt{\left\langle\left[I(t+\Delta t) - \langle I(t+\Delta t)\rangle\right]^2\right\rangle\left\langle\left[I(t) - \langle I(t)\rangle\right]^2\right\rangle}}, \quad (5)$$

where $I(t)$ is the output intensity of the chaotic laser, $<\bullet>$ designates the time average, and $\Delta t$ is the EOF delay time.

PE is first introduced by Bandt and Pompe to measure the complexity of time series [32,33]. For a given time series $X = \{x(1), x(2), ..., x(n)\}$, we reconstruct a m-dimensional vector $X_i = \{x(i), x(i+l), ..., x(i+(m-1)l)\}$, where m and l represent embedding dimension and embedding delay time respectively. Then the elements of $X_i$ is arranged in ascending order $x(i+(j_1-1)l) \leq x(i+(j_2-1)l) \leq ... \leq x(i+(j_m-1)l)$. The resulting permutations is $\{j_1, j_2...j_m\}$, which is one of the full array m!. The relative frequencies of the permutations are calculated as their probabilities $p_1, p_2...p_k$, k≤m!. The normalized PE is defined as:

$$H(p) = \frac{-\sum_{i=1}^{k} p_i \log p_i}{\log(m!)}. \quad (6)$$

The value of H(p) is between 0 and 1. Embedding dimension m is recommended between 3 and 7 in practice. Due to time constraints, we choose d=4 in this paper. Figure 2 shows the theoretical results of power spectrum, TDS, and complexity of chaotic laser with and without quantum noise injection. The chaotic laser operates at a bias current $J$=1.35$J_{th}$ and a feedback strength $\kappa$ = 5 $ns^{-1}$, corresponding to the experimental conditions. The bandwidth of the original chaos is about 2. 5 GHz according to the 80% bandwidth definition, as shown in Fig. 2(a1). The intensity of chaotic laser shows large amplitude fluctuations, which indicate that the laser evolves into the coherence collapse regime as shown in the inset of Fig. 2(a2). The quantum noise strength is defined using the quantum noise to ground noise strength ratio (QGSR) as follows:

$$QGSR = 5\log_{10}\left(\frac{Q}{G}\right) = 5\log_{10}(Q) - 45. \qquad (7)$$

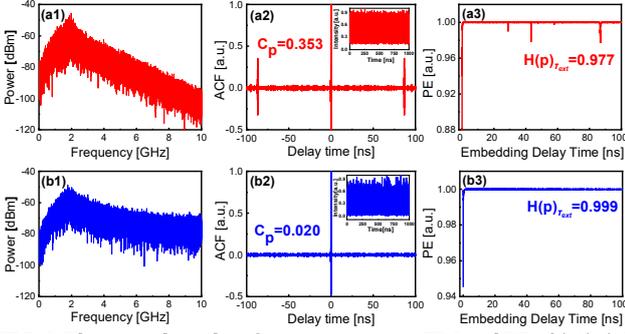

**FIG. 2.** Theoretical results of power spectrum, TDS and PE of (a1)-(a3) original chaotic laser and (b1)-(b3) chaos with quantum noise injection.

where Q represents the quantum noise intensity, and G is the ground noise intensity. In our experiment, the ground noise intensity G is 3.6mV that corresponds to the power level of -45dbm. The theoretical quantum noises with different bandwidth and strength are prepared by filtering and amplifying Gaussian white noise. With the bandwidth of 100 MHz and the strength of 16 dB quantum noise injection [Fig. 2(b1)-2(b3)], the power spectrum of the chaos is broadened and the TDS is effectively suppressed up to 94% compared to the original chaos. Meanwhile, the PE at $\tau_{ext}$ of chaotic laser also approaches to the maximum of 1 and the complexity of chaos is enhanced clearly.

Furthermore, the effects of quantum noise strength and bandwidth on the TDS and complexity of chaotic laser are investigated theoretically. Figures 3(a) and 3(b) show the $C_p$ and $H(p)_{\tau_{ext}}$ as a function of QGSR for various bandwidths of quantum noise. The $C_p$ of chaotic laser fall monotonically and the corresponding $H(p)_{\tau_{ext}}$ is enhanced quickly with increasing the QGSR. When the strength of quantum noise is strong enough, the minimum $C_p$ of 0.020 remains almost unchanged and the complexity approaches to the ideal value of 1. The TDS and complexity versus the quantum noise to chaos bandwidth ratio (QCBR) for various strengths of quantum noise are shown in Fig. 3 (c) and 3(d). At five QGSRs, the $C_p$ gradually reduces and $H(p)_{\tau_{ext}}$ monotonically enhances as the QCBR increases. The QCBR can be reached to 1:25 with the QGSR of 16 dB, in which case the $C_p$ is suppressed to the minimum and the $H(p)_{\tau_{ext}}$ is enhanced to the maximum.

We experimentally investigate the TDS suppression and complexity enhancement using quantum noise. The LD operates with a driven current J=1.3 $J_{th}$ and feedback strength η=3%, which outputs a chaotic laser. According to the 80% bandwidth definition, the bandwidth of the chaotic laser is about 2.5 GHz which is consistent with that of the theoretical simulation, as shown in Fig. 4(a1). Various bandwidth quantum noises are extracted by the homodyne detection, and the quantum noises are amplified and injected into the high-frequency modulation connector of the LD in Fig.1. We observe the time-frequency results of the chaotic laser with and without injecting quantum noise. Figure 4 shows the power spectrum, probability statistical distribution of laser intensity, TDS and complexity of original chaotic laser, quantum noise, and chaotic laser with injecting quantum noise. The TDS and complexity of the original chaotic laser are $C_p$=0.374 and $H(p)_{\tau_{ext}}$ = 0.983, and the skewness of its intensity distribution is 0.529, as shown in Fig. 4 (a1)-4(a2). The intensity distribution of the original chaotic laser deviates obviously from Gaussian random distribution. The quantum noise In Fig. 4(b1) has more than 14 dB clearance above the noise floor, which is achieved only by increasing the power to 4 mW. The intensity distribution of quantum noise obeys Gaussian random distribution and the skewness of the statistical distribution is 0.001, as shown in Fig. 4(b2). It is noted that with the injection of the bandwidth of 100 MHz and the strength of 16 dB quantum noise, the resulting TDS of the chaotic laser is suppressed from $C_p$=0.374 to $C_p$=0.023 and the corresponding complexity is enhanced from $H(p)_{\tau_{ext}}$ = 0.983 to $H(p)_{\tau_{ext}}$ = 0.999. as shown in Fig. 4(c2)-4(c3). Meanwhile, the probability statistical distribution of chaotic intensity is significantly improved and fits well with Gaussian random distribution. The experimental skewness of the intensity distribution is negligible with a value of 0.011. As expected, the chaotic TDS is significantly

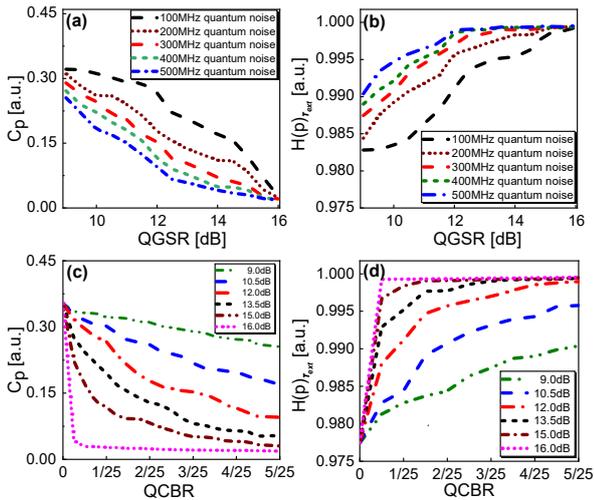

**FIG. 3.** Theoretical (a), (c) $Cp$ and (b), (d) $H(p)_{\tau_{ext}}$ of chaotic laser as functions of QGSR for five quantum noise bandwidths (a), (b): 100MHz, 200MHz, 300MHz, 400MHz, 500MHz and QCBR for five QGSRs (c), (d): 9.0dB, 10.5dB, 12.0 dB, 13.5 dB, 15.0 dB,16.0dB.

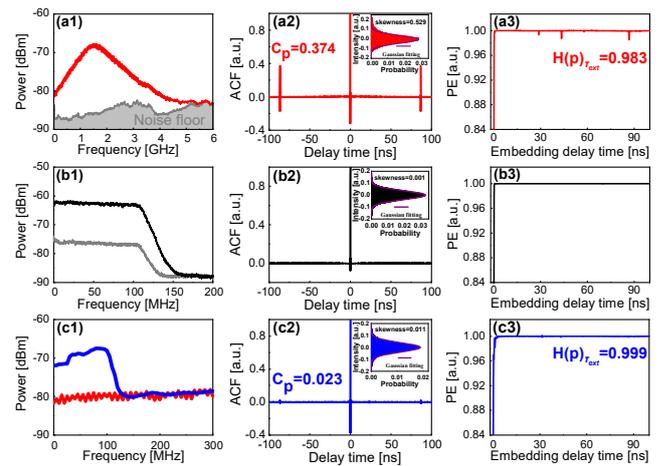

**FIG. 4.** Experimental results of power spectrum, TDS, probability statistical distribution of laser intensity, and complexity of (a1)-(a3) original chaotic laser, (b1)-(b3) quantum noise, and (c1)-(c3) chaotic laser with injecting quantum noise. The bandwidth and strength of quantum noise are 100 MHz and 16 dB.

suppressed and the complexity of the chaotic laser is effectively enhanced by using quantum noise.

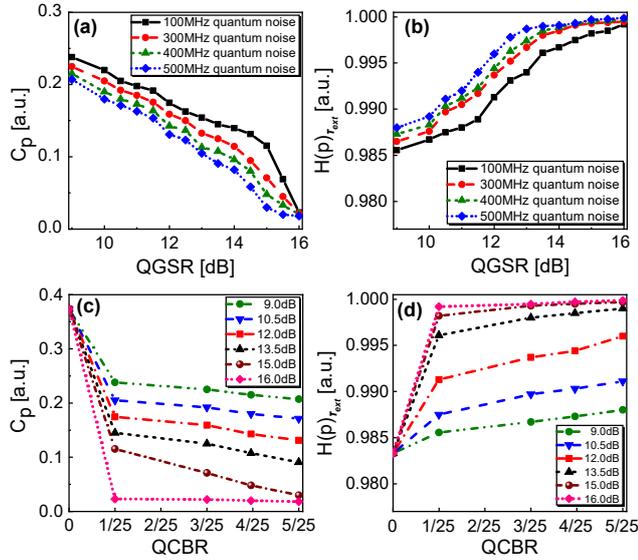

**FIG. 5.** Experimental (a), (c) $C_p$ and (b), (d) $H(p)_{\tau_{ext}}$ of chaotic laser as functions of QGSR at four quantum noise bandwidths (a), (b): 100MHz, 300MHz, 400MHz, 500MHz and QCBR at five QGSRs (c), (d): 9.0dB, 10.5dB, 12.0dB, 13.5 dB, 15.0 dB,16.0dB.

Moreover, we observe the influences of quantum noise strength and bandwidth on the TDS and complexity of the chaotic laser. Figures 5(a) and 5(b) show the $C_p$ and $H(p)_{\tau_{ext}}$ as a function of QGSR at four different quantum noise bandwidths: 100MHz, 300MHz, 400MHz, 500MHz. It is indicated that at all the four quantum noise bandwidths, the complexity is enhanced as the QGSR increases, which coincides with the TDS suppression. In Figs. 5(c) and 5(d) the $C_p$ and $H(p)_{\tau_{ext}}$ versus the QCBR for various quantum noise strengths are shown. When the QGSR (i.e. quantum noise strength) is fixed, the $C_p$ decreases and $H(p)_{\tau_{ext}}$ enhances monotonically as the QCBR increases. It is worth noting that the maximum $H(p)_{\tau_{ext}}$ and the minimum $C_p$ are simultaneously achieved with the 1:25 QCBR and 16 dB QGSR, which agrees well with the theory.

In conclusion, the effects of quantum noise bandwidth and strength on the chaotic TDS has been studied theoretically and experimentally. The results show that the TDS of chaos is suppressed up to 94% and the bandwidth suppression ratio of quantum noise to chaotic laser is 1:25 compared to the original chaos. Moreover, the TDS $C_p$ gradually reduces and the complexity $H(p)_{\tau_{ext}}$ monotonically enhances as the strength and bandwidth of quantum noise increase. Thus, narrow-bandwidth quantum noise substantially suppresses the TDS of wide-bandwidth chaotic laser, and improves the randomness of the chaotic signal. The technique is useful for understanding underlying randomness of chaotic lasers and leading to secure communication system with TDS suppression.

**Funding.** National Natural Science Foundation of China (61875147, 62075154, 61731014, 61961136002); Key R&D Program of Shanxi Province (International Cooperation, 201903D421049); Shanxi Scholarship Council of China (HGKY2019023,);

**Disclosures**. The authors declare no conflicts of interest.